\documentstyle[twocolumn,floats,ifthen,aps,prl]{revtex}

\begin{document} 
 
\draft 

\title{ 
Continuous quantum measurement with observer: 
pure wavefunction evolution instead of decoherence 
}

\author{Alexander N. Korotkov \footnote{Electronic mail: 
	akorotkov@ccmail.sunysb.edu}} 
\address{ 
GPEC, Departement de Physique, Facult\'e des Sciences de Luminy, 
Universit\'e de la M\'editerran\'ee, 
13288 Marseille, France \\ 
and Nuclear Physics Institute, Moscow State University,
Moscow 119899, Russia}
 
\date{\today} 
 
\maketitle 
 
\begin{abstract} 
        We consider a continuous measurement of a two-level 
system (double-dot) by weakly coupled detector (tunnel point contact 
nearby). While usual treatment leads to the gradual system 
decoherence  due to the measurement, we show 
that the knowledge of the measurement result can restore the pure
wavefunction at any time (this can be experimentally verified). 
The formalism allows to 
write a simple Langevin equation for the random evolution of the system
density matrix which is reflected and caused by the stochastic 
detector output. Gradual wavefunction ``collapse'' 
and quantum Zeno effect are naturally described by the equation.
\end{abstract} 
 
 
\narrowtext 
 
\vspace{1ex} 

        The problem of quantum measurements has a long history,
however, it still attracts considerable attention and causes
discussions and even some controversy, mainly about the wavefunction
``collapse'' postulate \cite{Neumann,Luders,Wheeler,Kempen}.  
One of the leading modern ideas is to
replace this postulate by the gradual decoherence  of the density
matrix due to the interaction with the detector, so that this
decoherence can be described by the Schr{\"o}dinger equation
and, hence, no additional postulate is necessary (see, e.g.\
Ref. \cite{Zurek}).
Let us also mention the ``hidden variables'' idea \cite{Bohm}
and the approach of a stochastic evolution of the wavefunction
\cite{Pearl,Gisin,Diosi,Wiseman,Belavkin,Hegerfeldt,Dalibard,Power,Plenio}. 
The renewed interest to the measurement problem 
is justified by the development of the experimental technique,
which allows more and more experimental studies of the quantum
measurement in optics and mesoscopic structures 
\cite{Itano,Aspect,Brune,Buks,Tittel}. 
The problem has also
close connection to the rapidly growing fields of quantum cryptography 
\cite{crypto} and quantum computing \cite{q-comp}.

        In the recent experiment \cite{Buks} with  ``which-path''
interferometer the suppression of Aharonov-Bohm interference 
due to the detection of which path an electron chooses, was observed. 
The weakly coupled quantum point contact was used as a detector.
The interference suppression in this experiment can be quantitatively 
described by the decoherence (dephasing \cite{Stern}) due to 
the measurement process 
\cite{Gurvitz,Aleiner,Levinson,Stodolsky} (see also Refs.\ 
\cite{Englert,Shnirman}). 

In the present paper we consider somewhat different setup which is simpler
and more basic in the context of quantum measurements:
two quantum dots occupied by one electron and a weakly coupled
detector measuring the position of the electron. As a detector
we assume small tunnel contact (``point contact'') close to the double-dot 
so that the detector barrier height depends on the electron position. 
This setup was analyzed in Ref.\ \cite{Gurvitz} (see also Ref.\
\cite{Stodolsky}) in which the equations for the double-dot density 
matrix evolution affected by the decoherence due to the measurement 
process, were derived. However, the decoherence approach cannot
describe the detector output that is a 
separate interesting problem \cite{Stodolsky,Gur-98} analyzed
in the present paper. We answer two related questions: 
how the detector current looks like (as a function of time) 
and what is the proper double-dot density matrix 
for particular detector output.
(Notice that our result for the first question 
contradicts the point of view presented
in Ref.\ \cite{Stodolsky}.)

        We show that the decoherence rate derived in Refs. 
\cite{Gurvitz,Aleiner,Levinson,Stodolsky} coincides with the lower
bound determined by the knowledge about the system 
gradually acquired during the continuous measurement (thus proving 
that the considered model of point contact corresponds to an  
``ideal'' detector). 
This lower bound is derived assuming that the system can be 
still described by the pure wavefunction after each particular 
realization of the random detector output. Hence, the fact 
that the lower bound is really achieved for the point contact
as a detector, leads us to 
the conclusion that the decoherence in this case 
is just a consequence of ignoring the measurement 
result, i.e. averaging over all possible realizations. 
The observer who follows the detector output is able to
obtain the complete knowledge about the system: he 
knows the wavefunction at each moment of time. 

The measurement process modifies the wavefunction, for example,
leading to gradual localization. From the observer's
point of view the evolution of the wavefunction can be
described as a stochastic process related to the detector
output. We develop a simple formalism of this evolution 
and briefly discuss the philosophical aspect of the presented
result. The formalism can be applied to more general case
of a two-level system measured by weakly coupled
detector; however, for the definiteness we speak about
the double-dot and point contact.

\vspace{0.3cm}

        Similar to Ref.\ \cite{Gurvitz} let us describe
the double-dot system and the measuring point contact by the Hamiltonian
        \begin{equation}
        {\cal H}={\cal H}_{DD}+{\cal H}_{PC}+{\cal H}_{int},
        \label{hamilt}  \end{equation}
where 
        \begin{equation}
{\cal H}_{DD} = \frac{\epsilon }{2} (c_1^\dagger c_1-c_2^\dagger c_2)+ 
H (c_1^\dagger c_2+ c_2^\dagger c_1) 
        \label{ham_DD}\end{equation}
is the standard Hamiltonian of the double-dot system \cite{shift} 
with tunneling coupling $H$ between dots ($H$ is assumed to be real),
        \begin{equation}
{\cal H}_{PC}=\sum_l E_la_l^\dagger a_l +\sum_r E_r a_r^\dagger a_r +
\sum_{l,r} T (a_r^\dagger a_l+a_l^\dagger a_r)
        \label{ham_PC}\end{equation}
describes the tunneling through the point contact (for simplicity
$T$ is real and does not depend on energies), and the coupling 
between the double-dot and the detector is assumed to be 
        \begin{equation}
{\cal H}_{int}= \sum_{l,r} \Delta T \, c_2^\dagger c_2 
(a_r^\dagger a_l + a_l^\dagger a_r),
        \label{ham_int}\end{equation}
i.e.\ the tunneling matrix element for the point contact is $T$ 
when the first dot is occupied while it is $T+\Delta T$ when
the electron is in the second dot. The voltage $V$ across the point
contact is sufficiently large, $eV\gg T^2\rho$ ($\rho$ is the density
of states), so that the simple description of the point contact 
is possible (see Ref.\ \cite{Gurvitz}). Basically we can say that
the average current $I_1=2\pi T^2 \rho_l \rho_r e^2V/\hbar$ flows through 
the detector when the electron is in the first dot, and the current is
$I_2=I_1+\Delta I=2\pi (T+\Delta T)^2\rho_l\rho_r e^2V/\hbar$ 
when the second dot is occupied.

        We make an important assumption of weak coupling between
the double-dot and the detector (actually, it would be better
to call it ''weakly responding'' detector),
        \begin{equation}
| \Delta I | \ll I_0= (I_1 +I_2)/2,
        \label{weak}\end{equation}
so that many electrons ($N \agt (I_0/\Delta I)^2$) should pass through 
the point contact before
the observer is able to distinguish which dot is occupied
(i.e.\ when the uncertainty due to the detector shot noise becomes
less than $\Delta I$). 
This assumption is necessary to allow the classical description 
of the detector, namely to neglect the coherence between the 
quantum states with different number of electrons passed through
the detector (we implicitly assume that the corresponding ``collapse''
happens on the time scale $t \ll (e/I_0) (I_0/\Delta I)^2$, much faster
than typical evolution of the double-dot density matrix).

        One of the main results of Ref.\ \cite{Gurvitz} is
the equation for the decoherence rate $\Gamma_d$ of the nondiagonal 
element $\sigma_{12}(t)$ of the double-dot density matrix due to 
the measurement by the point contact:
$\Gamma_d = (\sqrt{I_1/e}-\sqrt{I_2/e})^2/2$.
In the weakly-coupled limit (\ref{weak}) it can be replaced
by 
        \begin{equation}
\Gamma_d= \frac{1}{8} \frac{(\Delta I)^2}{eI_0}.
        \label{gam_d}\end{equation}    
The decoherence has an obvious relation to the low frequency 
shot noise in the detector (the origin can be traced to Eq.\ 
(\ref{ham_int})), so let us write Eq.\ (\ref{gam_d}) in the form 
        \begin{equation}
\Gamma_d = \frac{1}{4} \frac{(\Delta I)^2}{S_I},
        \label{gam_d-m}\end{equation} 
where $S_I=2eI_0$ is the usual Schottky formula for the shot noise
spectral density $S_I$. 
Equation (\ref{gam_d-m}) has been also obtained in Refs.\
\cite{Aleiner,Levinson} for the quantum point contact as a detector,
the difference in that case is $S_I=2eI_0 (1-{\cal T})$ 
where ${\cal T}$ is the transparency of the channel \cite{Lesovik} 
(while in the case considered above we implicitly assumed  
${\cal T}\ll 1$ 
\cite{largeT}). As shown in Ref.\ \cite{Stodolsky}, Eq.\ (\ref{gam_d-m})
should be modified (decoherence rate increases) if the phase of 
transmitted and reflected electrons
in the detector is sensitive to the double-dot state; we assume that 
there is no such a dependence in our case. 

        Concluding the introductory part of the paper let us 
write the full equation for the double-dot density matrix
in the decoherence approach: 
        \begin{eqnarray}
{\dot \sigma}_{11} &=& -{\dot \sigma}_{22}= 
\frac{iH}{\hbar} (\sigma_{12}-\sigma_{21}), 
        \label{s_11}    \\
{\dot \sigma}_{12} &=& \frac{i\epsilon }{\hbar} \sigma_{12} + 
        \frac{iH}{\hbar} (\sigma_{11}-\sigma_{22}) -
         \frac{1}{4} \frac{(\Delta I)^2}{S_I} \sigma_{12}.
        \label{s_12}\end{eqnarray}

        Notice that the decoherence rate (\ref{gam_d-m}) was derived 
in Refs.\ \cite{Gurvitz,Aleiner,Levinson,Stodolsky} without any account of
the information provided by the detector, implicitly assuming that
the measurement result is just ignored. Now let us study how
this additional information affects the double-dot density matrix.

\vspace{0.3cm}

        We start with the completely classical case when there is no 
tunneling between dots ($H=0$) and
the initial density matrix of the system does not have nondiagonal 
elements, $\sigma_{12}(0)=0$ (then obviously $\sigma_{12}(t)=0$ 
for any $t>0$). 
We can assume that the electron is actually located in one of the 
dots, but we just do not know exactly in which one, and that is why we 
use probabilities $\sigma_{11}(0)$ and $\sigma_{22}(0)=1-\sigma_{11}(0)$. 
The detector output is the fluctuating current $I(t)$. 
The fluctuations grow when we examine $I(t)$
at smaller time scales, so we need some averaging in time (``low-pass
filtering''), at least in order to neglect the problem of individual electrons
passing through the point contact. Let us always work 
at sufficiently low frequencies, $f\ll S_I/e^2$, for which 
the low frequency limit $S_I$ for the spectral density is well achieved.
        
        Provided that $i$th dot is occupied, the probability 
to have a particular value for the current averaged over time $\tau$,  
$\langle I\rangle =\int_0^\tau I(t) dt$, 
 is given by the Gaussian distribution 
        \begin{eqnarray}
P_i(\langle I\rangle , \tau ) &=& (2\pi D)^{-1/2} 
\exp \left( -(\langle I\rangle -I_i)^2/2D \right) , 
        \nonumber \\
 D &=& S_I/2\tau .
        \label{Gauss}\end{eqnarray}
Notice that this equation obviously does not change if we divide the 
time interval $\tau$ into pieces and integrate over all
possible average currents for each piece
(to consider only positive currents the typical 
timescale $\tau$ should be sufficiently long, $S_I/\tau \ll I_i^2$,
that is always satisfied within the assumed low frequency range).
After the measurement during
time $\tau$ the observer acquires additional knowledge about the system
and should change the probabilities $\sigma_{ii}$ according to the standard
Bayes formula. (It says that a posteriori probability $p'(A)$ of an event 
$A$ after the knowledge that the event $F$ has happened, 
is equal to $p'(A)=p(A)p(F|A)/\sum_B [p(B)p(F|B)]$ where $p(A)$ is  
a priori probability and $p(F|A)$ is the conditional probability of 
event $F$ given event $A$.) Hence,
        \begin{eqnarray}
\sigma_{11} (\tau ) &&= \sigma_{11}(0) \exp [ 
-(\langle I\rangle -I_1)^2/2D ] 
        \nonumber \\
&&\, \times \left\{ \sigma_{11}(0)\exp [ -(\langle 
 I\rangle -I_1)^2/2D ] \right.  
        \nonumber \\
 && \,\,\,\, + \left. \sigma_{22}(0) \exp  [- (\langle I\rangle -I_2)^2/2D] 
\right\} ^{-1} ,
        \nonumber \\
\sigma_{22} (\tau ) &&= 1-\sigma_{11} (\tau ) .
        \label{s1-2}\end{eqnarray}
Notice that we do not use any ``collapse'' postulate here because
we speak so far about the classical measurement.

\vspace{0.3cm}

        Now let us assume that the initial state was fully coherent,
$\sigma_{12}(0)=\sqrt{\sigma_{11}(0)\sigma_{22}(0)}$ (while still 
$H=\epsilon =0$). Since the detector is sensitive only 
to the position of electron, the detector current 
will behave exactly the same way \cite{obviously} 
and the probability of a particular
value $\langle I\rangle$ is still given by 
        \begin{equation}
P(\langle I\rangle , \tau)=\sigma_{11}(0) P_1 (\langle I\rangle ,\tau )+ 
\sigma_{22}(0) P_2(\langle I\rangle ,\tau ).
        \label{prob}\end{equation}
        After  the  measurement  during  time  $\tau$   we   should 
obviously 
assign the same values for $\sigma_{11}(\tau )$ and $\sigma_{22}(\tau )$ 
as in Eq.\ (\ref{s1-2}), but the question is not so trivial for the
nondiagonal element $\sigma_{12}(\tau )$. Nevertheless, we can easily 
write the upper bound: 
        \begin{equation}
\mbox{Re} \, \sigma_{12} (\tau ) \leq |\sigma_{12}(\tau )| \leq 
\sqrt{ \sigma_{11}(\tau ) \sigma_{22}(\tau )}. 
        \label{u-b}\end{equation}

        Let us imagine the observer who does not want to know
the result of the measurement (which actually exists!). Then using
the probability distribution of different outcomes given by
Eq.\ (\ref{prob}) and the upper bound (\ref{u-b}) for each
realization, he can calculate the upper bound for $\sigma_{12}$
(disregarding the actual result):
        \begin{eqnarray}
\langle \mbox{Re} \, \sigma_{12}(\tau )\rangle \leq 
\int \sqrt{\sigma_{11}(\tau )\sigma_{22}(\tau )} \, 
        P(\langle I\rangle ,\tau ) \, d\langle I\rangle 
        \nonumber \\
= \sqrt{\sigma_{11}(0)\sigma_{22}(0)} \, 
\exp \left( -\frac{(\Delta I)^{2}\tau }{4S_I} \right) .
        \end{eqnarray}
This upper bound exactly coincides with the actual result
given by decoherence approach (\ref{s_12}). This fact
forces us to accept somewhat surprising statement
that Eq.\ (\ref{u-b}) gives not only the upper bound, but 
the true value of the nondiagonal 
matrix element, i.e.\ the pure state remains pure  
after the measurement (no decoherence occurs) if we know the
measurement result \cite{strong}.
        
        Simultaneously, we prove that the point contact detector 
considered in Refs.\ \cite{Gurvitz,Aleiner,Levinson} causes the slowest
possible decoherence of the measured system (disregarding the measurement 
result), and hence represents an ideal detector in this sense. 
In contrast, the result 
of Ref.\ \cite{Shnirman} shows that a single-electron transistor with 
large tunnel resistances and biased by relatively large voltage, 
is not an ideal detector (for the same amount
of the back-influence on the system it provides an observer with 
less information than an ideal detector). Similarly, the generalization 
of the quantum point contact considered in Ref.\ \cite{Stodolsky} 
describes a non-ideal detector.

        If the initial state of the double-dot is not purely coherent, 
$|\sigma_{12}(0)|<\sqrt{\sigma_{11}(0)\sigma_{22}(0)}$, 
we can treat it as the statistical combination of purely 
coherent and purely incoherent states with the same
$\sigma_{11}(0)$ and $\sigma_{22}(0)$. Then instead
of Eq.\ (\ref{u-b}) we have
        \begin{equation}
\sigma_{12}(\tau )= \sigma_{12}(0) \,
\frac{\left[ \sigma_{11}(\tau ) \sigma_{22}(\tau ) \right] ^{1/2}}
{\left[ \sigma_{11}(0) \sigma_{22}(0) \right] ^{1/2}}. 
        \label{s_12-g}\end{equation}
Eq.\ (\ref{s_12-g}) together with Eq.\ (\ref{s1-2}) is the central
result of the present paper; these equations give the density matrix 
of the measured system with account of the measurement result
\cite{alternat}.

\vspace{0.3cm}

        The measurement should lead to the localization of the
wavefunction in one of the dots. This is a random process, and
the observer who continuously follows the detector output can 
describe it as the random evolution of the wavefunction (provided
the pure initial state), or more generally the random evolution 
of the density matrix.
        Eqs.\ (\ref{Gauss})--(\ref{prob}) and (\ref{s_12-g})
allow to simulate this evolution. For example, we can use
Monte-Carlo method and do the following. First we choose
the timestep $\tau$ satisfying inequalities $e^2/S_I \ll \tau
\ll S_I/(\Delta I)^2$ and draw a random number for 
$\langle I\rangle$ according to the distribution (\ref{prob}). 
Then we update $\sigma_{11}(t)$ and $\sigma_{22}(t)$ using
this value of $\langle I\rangle$ and repeat the procedure many
times (the distribution for the current averaged over 
the interval $\Delta t=\tau$ 
is new every timestep because of changing $\sigma_{ii}(t)$ which
are used in Eq.\ (\ref{prob})).
The nondiagonal matrix element can be calculated at any time
using Eq.\ (\ref{s_12-g}).

        This Monte-Carlo procedure can be equivalently described by 
the nonlinear Langevin-type equation for the density matrix evolution 
(equation for $\sigma_{11}$ is sufficient):
        \begin{equation}
{\dot \sigma}_{11}={\cal R}=
-\sigma_{11}\sigma_{22}\, \frac{2\Delta I}{S_I}
\left[\frac{\sigma_{22}-\sigma_{11}}{2}\, \Delta I +\xi (t)  \right] ,
        \label{s_11-evol}\end{equation}
where the random process $\xi (t)$ has zero average and the spectral density
$S_\xi = S_I$ (only low-frequency limit matters). The term in square
brackets is equal to $I(t)-I_0$, so it is directly related to the
detector output.
        One can easily check that calculation of actual 
$\sigma_{11}(t)$ evolution for known detector output $I(t)$ using
Eq.\ (\ref{s_11-evol}) coincides with the direct result given by 
Eq.\ (\ref{s1-2}).

        Equation (\ref{s_11-evol}) is closely connected with the 
Quantum State Diffusion approach of Refs.\ 
\cite{Gisin,Diosi,Wiseman,Belavkin} (for review, see Ref.\ \cite{Plenio}). 
Actually, it is possible to 
obtain mathematically such a stochastic differential equation for 
any equation for the density operator \cite{Gisin,Diosi,Wiseman}. 
In our treatment, however, we derived Eq.\ (\ref{s_11-evol})
using only basic physical reasoning. 

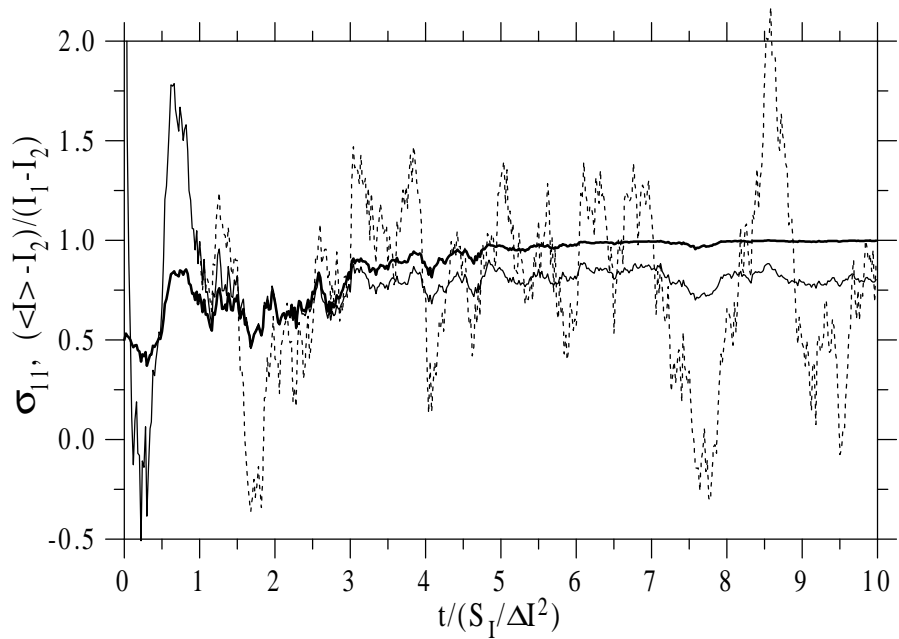
\begin{figure}
\caption{Thick line:  particular Monte-Carlo realization 
of $\sigma_{11}$ evolution in time during the measurement of uncoupled
dots, $H=0$. 
Initial state is symmetric, $\sigma_{11}(0)=\sigma_{22}(0)=1/2$,
while the measurement leads to gradual localization. 
Initially pure wavefunction remains pure at any time $t$. 
Thin line shows the corresponding detector current 
$\langle I \rangle$ averaged over the whole time interval starting 
from $t=0$ while the dashed line is the current averaged over
the running window with duration $S_I/(\Delta I)^2$.
 }
\label{fig1}\end{figure}

        Figure \ref{fig1} shows a particular result of the Monte-Carlo
simulation for the symmetric initial state, $\sigma_{11}(0)=
\sigma_{22}(0)=1/2$ (notice that $\sigma_{12}(0)$ does not
affect the evolution if $H=0$). Thick line shows the random evolution 
of $\sigma_{11}(t)$. Equation (\ref{s_11-evol}) describes the gradual 
localization in one of the dots (first dot in case of Fig.\ \ref{fig1}).
Let us define the typical localization time $\tau_{loc}$ as
$\tau_{loc}=2S_I/(\Delta I)^2 $ (we choose the exponential factor
at $\sigma_{11}=\sigma_{22}=1/2$). Then it is exactly equal to the time
$\tau_{dis}=2S_I/(\Delta I)^2$ necessary for the observer to distinguish
between two states (defined as the relative shift of two Gaussians by two 
standard deviations), and $\tau_{loc}=\tau_d/2$ where $\tau_d=\Gamma_d^{-1}$.
The probability of final localization in the first dot
is equal to $\sigma_{11}(0)$ (as it should be) that can be easily proven 
because the procedure described above conserves $\sigma_{11}(\tau )-
\sigma_{22}(\tau )$ averaged over realizations.
 The detector current 
basically follows the evolution of $\sigma_{11}(t)$ but the additional noise
is large and depends on the bandwidth. The dashed line in Fig.\ 
\ref{fig1} shows the detector current averaged over the ``running window''
with the duration $\Delta t= S_I/(\Delta I)^2$ while the thin solid
line is current $\langle I\rangle$ averaged starting from $t=0$. 

        Our result for the detector current contradicts the 
statement made in Ref.\ \cite{Stodolsky} that the detector output
in each particular realization should correspond to the average 
double-dot population, $\langle I\rangle \simeq \sum_i I_i\sigma_{ii}(0)$,
which we believe is incorrect as well as the statement that 
$\sigma_{ii}$ can be measured in a single experiment ``without
a collapse of wavefunction''.

\vspace{0.3cm}

        Now let us consider the general case of the double-dot
system with non-zero tunneling $H$ between dots. 
If the frequency $\Omega$ of ``internal''
oscillations in the double-dot is sufficiently low so that 
the low-frequency limit for the detector shot noise is well achieved,
        \begin{equation}
\Omega = (4H^2+\epsilon^2)^{1/2}/\hbar \ll S_I/e^2 ,
        \end{equation}
then we can use the same formalism just adding the slow 
evolution due to finite $H$ (the product $\Omega \tau_{loc}$
can be both larger or smaller than unity, so in this sense
the coupling between double-dot and the detector can be
arbitrary large). 
The particular realization can be
either simulated by Monte-Carlo procedure similar to that outlined 
above [now update of $\sigma_{12}(t)$ using Eq.\ (\ref{s_12-g}) should
be necessarily done at each timestep] 
 or equivalently described by the corresponding coupled 
Langevin equations which are the counterpart of Eqs.\ 
(\ref{s_11})--(\ref{s_12}): 
        \begin{eqnarray}
{\dot \sigma}_{11} &=& -{\dot \sigma}_{22}=\frac{-2H}{\hbar} \, 
        \mbox{Im}(\sigma_{12}) +{\cal R},
        \label{s11-g}   \\
{\dot \sigma}_{12} &=& \frac{i\epsilon }{\hbar} \sigma_{12} 
+\frac{iH}{\hbar}(\sigma_{11}-\sigma_{22}) +
\frac{\sigma_{22}-\sigma_{11}}{2\sigma_{11}\sigma_{22}}\, {\cal R} \sigma_{12} 
        \nonumber \\
&& -\gamma_d\sigma_{12},
        \label{s12-g}\end{eqnarray}
where ${\cal R}$ is given by Eq.\ (\ref{s_11-evol}) and the last term 
in Eq.\ (\ref{s12-g}) will be discussed later ($\gamma_d =0$ for an
ideal detector). 

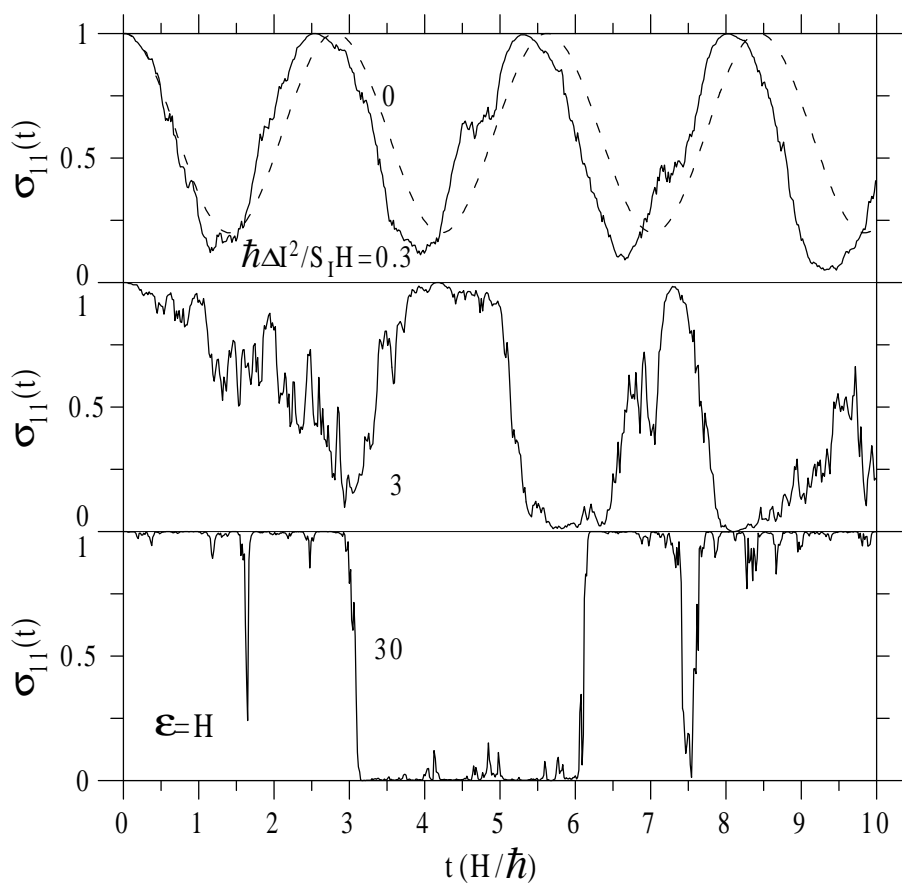
\begin{figure}
\caption{
Random evolution of $\sigma_{11}$ (particular Monte-Carlo realizations)
for asymmetric double-dot, $\epsilon =H$, with the electron initially
in the first dot, $\sigma_{11}(0)=1$, for different 
strength of coupling with detector: 
${\cal C}=\hbar (\Delta I)^2/S_IH=$0.3, 3, and 30
from top to bottom. Dashed line represents ${\cal C}=0$ (unmeasured
double-dot). Increasing coupling with detector destroys the quantum
oscillations (while wavefunction remains pure at any $t$), slows down
the transitions between states (Quantum Zeno effect), and for 
${\cal C}\gg 1$ leads to uncorrelated jumps between well localized
states.
 }
\label{fig2}\end{figure}

        Figure \ref{fig2} shows particular results of the Monte-Carlo
simulations for the double-dot with $\epsilon =H$ and different
strength of the interaction with an ideal detector. The electron is initially
located in the first dot, $\sigma_{11}(0)=1$. The dashed line shows
the evolution of $\sigma_{11}$ with no detector. Notice that because
of the energy asymmetry, the initial asymmetry of the electron location 
remains in this case for infinite time. When the interaction with detector, 
${\cal C} =\hbar(\Delta I)^2/S_IH$, is relatively
small (top solid line), the evolution of $\sigma_{11}$ is close to
that without the detector. However, the electron gradually ``forgets''
the initial asymmetry and the evolution can be described as the slow
variation of the phase and amplitude of oscillations (recall that
the wavefunction remains pure). In the decoherence approach 
(averaging over realizations) this corresponds to $\sigma_{11}\rightarrow 
1/2$ at $t\rightarrow \infty$ \cite{Gurvitz}.

        When the coupling with the detector increases, the 
evolution significantly changes (middle and bottom curves in Fig.\
\ref{fig2}). First, the transition between dots slows down
(Quantum Zeno effect \cite{Misra}; see also Refs.\ 
\cite{Gisin,Belavkin,Power,Itano,Gurvitz,Shnirman,Gur-98}).
Second, while the frequency of transitions decreases with increasing
interaction with detector (at sufficiently strong coupling), 
the time of a transition (sort of ``traversal'' time) decreases,
so eventually we can say about uncorrelated ``quantum jumps'' 
between states. The case ${\cal C} \gg 1$ is completely analogous
to the standard description of the quantum Zeno effect with
frequent wavefunction reductions.

        In a regime of small coupling with detector, ${\cal C}\ll 1$, 
the detector output is too noisy to follow the
evolution of $\sigma_{ii}$. It does not give an accurate information  
about the electron position and, correspondingly, only slightly 
affects the oscillations. On contrary,
when ${\cal C}\gg 1$ the detector accurately informs about the
position of electron and the jumps between states, and simultaneously
destroys the internal oscillating dynamics of the system.  

        Equations (\ref{s11-g})--(\ref{s12-g}) can be generalized
for a nonideal detector, $\Gamma_d > (\Delta I)^2/4S_I$ 
(as in Ref.\ \cite{Shnirman}), 
which gives an observer less information than possible in principle. 
Let us model a nonideal detector
as two ideal detectors ``in parallel'', so that observer can read
the output of the first of them while the output of the second
detector is disregarded. Then the information loss can be represented
by the extra decoherence term $-\gamma_d \sigma_{12}$ in Eq.\
(\ref{s12-g}) where $\gamma_d=\Gamma_d-(\Delta I)^2/4S_I$. 
The limiting case of a nonideal detector is the detector 
with no output (just an environment) or with disregarded output.
Then the evolution equations reduce to the standard decoherence
case described by Eqs. (\ref{s_11})--(\ref{s_12}).

        For nonideal detector it is meaningful to keep our old 
definition of the localization time, $\tau_{loc}=\tau_{dis}=
2S_I/(\Delta I)^2$ while decoherence (in decoherence approach) 
occurs faster, $\tau_d<2\tau_{loc}$. Actually, this means that
if another observer is able to get more information (to read
the output of the second detector in a model above), then
for him the localization time will be shorter. In other words,
we define localization time not as a real physical quantity  
(that is meaningless because observer cannot check it) but
as a quantity related to observer's information. Similarly,
we can define the effective decoherence time as 
$\tau_d'=\gamma_d^{-1}$.

\vspace{0.3cm}

        The main point of the present paper is that 
Eqs.\ (\ref{s11-g})--(\ref{s12-g}) can be used not only to simulate the
measurement process, but also to obtain  
the actual evolution of the density matrix in an experiment provided
the known detector output $I(t)$ (high-frequency component of the
output can be suppressed) and initial condition $\sigma_{ij}(0)$.
For this purpose the term ${\cal R}$ given by Eq.\ (\ref{s_11-evol})
should be replaced by 
        \begin{equation}
{\cal R}=-\sigma_{11}\sigma_{22}\, \frac{2\Delta I}{S_I} 
\left[ I(t)-I_0 \right] .
        \end{equation}
        Notice that even if the initial state is completely
random, $\sigma_{11}=\sigma_{22}=1/2$, $\sigma_{12}=0$, the
nondiagonal matrix element appears during the measurement 
because of acquired information, so that sufficiently long 
observation with an ideal detector leads to almost pure
wavefunction (of course, this wavefunction does not have
direct relation to the initial state but emerges during
the measurement).

\vspace{0.3cm}

        Let us briefly discuss the philosophical aspect of the developed 
formalism. The statement that the pure wavefunction remains pure
during the continuous  measurement by an ideal detector (with  known  
result)  may  seem surprising 
at first, however, we easily recognize that this is a direct analogy
to the ``orthodox'' situation of a ``sharp'' measurement (the wavefunction
is pure after the ``collapse''). Another important point
is that the density matrix is in some sense observer-dependent.
        If an observer disregards the detector output, he can
either average over all possible detector outcomes or assume
the decoherence; both ways give the same result. Now if two
observers have different level of access to the detector information
(as, for example, in the model of nonideal detector considered above),
then the density matrix for them will be different. Nevertheless,
the observer with less information can safely use his density matrix
for all purposes; the only difference -- he will be able to make  
less accurate predictions than the observer with complete knowledge 
of the detector output.
There is no sense to speak about ``actual'' density matrix, 
it is meaningful to speak only about ``accessible'' density matrix.
This statement obviously contradicts the point of view that the density
matrix represents the objective reality. Simultaneously, this statement
is completely consistent with the ``orthodox'' (Copenhagen) point of 
view that in quantum mechanics the reality is closely related to our 
knowledge about it, so the density matrix represents the indivisible
mixture of the reality and our information about it.

        If the knowledge of the detector output is not used in the 
experiment, then the post-measurement density matrices 
should be averaged, leading (equivalently) to decoherence. 
On contrary, one can devise 
an experiment in which the subsequent system evolution depends on
the preceding measurement result; then the only proper description
is the pure wavefunction (for simplicity we assume ideal detector).

For example, let us consider the double-dot with $H=0$
and fully coherent symmetric initial state. According to our formalism,
after the measurement during some time $\tau$ (most interesting
case is $\tau \alt \tau_{loc}$) the wavefunction remains pure
but becomes asymmetric (Eqs.\ (\ref{s1-2}) and (\ref{s_12-g})). 
This means that if an experimentalist can switch off the detector
at $t=\tau $, 
reduce the barrier between the dots (create finite $H$) 
and change the relative energies of the dots in a proper way, then
after some definite time period the electron can be moved to the
first dot with the probability equal to unity (the corresponding
parameters can be easily calculated using $\sigma_{ij}(\tau )$
\cite{move})
that can be checked by the detector switched on again.
Alternatively, using the knowledge of $\sigma_{ij}(\tau )$ 
an experimentalist can exactly prepare the ground state of the coupled
double-dot system and check it, for example, by the photon
absorption.
        Another experimental idea is to start with completely random
state of the double-dot with finite $H$ and then gradually
(most interesting case is $\Omega \tau_{loc}\alt 1$)
obtain almost pure wavefunction using the detector output $I(t)$
and Eqs.\ (\ref{s11-g})--(\ref{s12-g}). The final test of
the wavefunction is similar to that considered above. 

An experiment of this kind would be able to verify the formalism 
developed in the present paper. While such an experiment is
still a challenge for the present-day technology, we can
hope that it will become realizable in the nearest future.

        In conclusion, we developed a simple formalism 
for the evolution of double-dot density matrix with account
of the result of the continuous measurement by weakly
coupled (weakly responding) point contact. The formalism 
is suitable for any two-level system measured by  
weakly coupled detector.  

        The author thanks S. A. Gurvitz, D. V. Averin, and K. K. Likharev 
for fruitful discussions. The work was supported in 
part by French MENRT (PAST), Russian RFBR, and 
Russian Program on Nanoelectronics. 

The author will be grateful to receive comments.

\begin{references} 

\bibitem{Neumann} J. von Neumann, {\it Mathematical Foundations of
        Quantum Mechanics} (Princeton Univ. Press, Princeton, NJ, 1955).

\bibitem{Luders} G. L\"uders, Ann. Phys. (Leipzig) {\bf 8}, 323 (1951).

\bibitem{Wheeler} Quantum Theory of Measurement, ed. by
        J. A. Wheeler and W. H. Zurek (Princeton Univ. Press, 
        Princeton, NJ, 1983).
        
\bibitem{Kempen} N. G. van Kempen, Physica A {\bf 153}, 97 (1988).

\bibitem{Zurek} W. H. Zurek, Phys. Today, {\bf 44} (10), 36 (1991);
        Phys. Rev. D {\bf 24}, 1516 (1981).

\bibitem{Bohm} D. Bohm and J. Bub, Rev. Mod. Phys. {\bf 38}, 453 (1966).

\bibitem{Pearl} P. Pearl, J. Stat. Phys. {\bf 41}, 719 (1985).

\bibitem{Gisin} N. Gisin, Phys. Rev. Lett. {\bf 19}, 1657 (1984);
        N. Gisin and I. C. Percival, J. Phys. A {\bf 25}, 5677 (1992).

\bibitem{Diosi} L. Diosi, J. Phys. A {\bf 21}, 2885 (1988).

\bibitem{Wiseman} H. M. Wiseman and G. J. Milburn, Phys. Rev. A {\bf 47},
        1652 (1993).

\bibitem{Belavkin} V. P. Belavkin and P. Staszewsky, Phys. Rev. A {\bf 45},
        1347 (1992).

\bibitem{Hegerfeldt} G. C. Hegerfeldt, Phys. Rev. A {\bf 47}, 449 (1993). 

\bibitem{Dalibard} J. Dalibard, Y. Castin, and K. Molmer, 
        Phys. Rev. Lett. {\bf 68}, 580 (1992).

\bibitem{Power} W. L. Power and P. L. Knight, Phys. Rev. A {\bf 53},
        1052 (1996).

\bibitem{Plenio} M. B. Plenio and P. L. Knight, Rev. Mod. Phys. 
	{\bf 70}, 101 (1998).

\bibitem{Itano} W. M. Itano, D. J. Heinzen, J. J. Bollinger,
        and D. J. Wineland, Phys. Rev. A {\bf 41}, 2295 (1990).

\bibitem{Aspect} A. Aspect, J. Dalibar, and G Roger, Phys. Rev. Lett.
        {\bf 49}, 1804 (1982).

\bibitem{Brune} M. Brune, E. Hagley, J. Dreyer, X. Maitre, A. Maali,
	C. Wunderlich, J. M. Raimond, and S. Haroche,
	 Phys. Rev. Lett. {\bf 77}, 4887 (1996).

\bibitem{Buks} E. Buks, R. Schuster, M. Heiblum, D. Mahalu,
        and V. Umansky, Nature {\bf 391}, 871 (1998).

\bibitem{Tittel} W. Tittel, J. Brendel, H. Zbinden, and N. Gisin,
	quant-ph/9806043.

\bibitem{crypto} C. H. Bennett, G. Brassard, and N. D. Mermin,
        Phys. Rev. Lett. {\bf 68}, 557 (1992);
        A. K. Ekert, Phys. Rev. Lett. {\bf 67}, 661 (1991).

\bibitem{q-comp} C. Bennett, Phys. Today, Oct. 1995, 24 (1995);
        P. Shor, SIAM J. Computing {\bf 26}, 1486 (1997);
        A. Ekert and R. Jozsa, Rev. Mod. Phys. {\bf 68}, 733 (1996).

\bibitem{Stern} A. Stern, Y. Aharonov, and J. Imry,
        Phys. Rev. A {\bf 41}, 3436 (1990).

\bibitem{Gurvitz} S. A. Gurvitz, Phys. Rev. B {\bf 56}, 15215 (1997).

\bibitem{Aleiner} I. L. Aleiner, N. S. Wingreen, and Y. Meir,
        Phys. Rev. Lett. {\bf 79}, 3740 (1997).

\bibitem{Levinson} Y. Levinson, Europhys. Lett. {\bf 39}, 299 (1997).

\bibitem{Stodolsky} L. Stodolsky, quant-ph/9805081.

\bibitem{Englert} B.-G. Englert, Phys. Rev. Lett. {\bf 77}, 2154 (1996).

\bibitem{Shnirman} A. Shnirman and G. Sch\"on, 
	Phys. Rev. B {\bf 57}, 15400 (1998). 

\bibitem{Gur-98} S. A. Gurvitz, quant-ph/9806050.

\bibitem{shift} The energy shift due to interaction with the detector 
        discussed in Ref.\ \cite{Stodolsky} can be included in $\epsilon$.
        (Actually, the shift is zero in our particular model because
        the phase of the detector electrons is not altered.)

\bibitem{Lesovik} G. B. Lesovik, JETP Lett. {\bf 49}, 591 (1989). 

\bibitem{largeT} In the case $1-{\cal T}\ll 1$ Eq.\ (\protect\ref{weak})
        should be replaced by $|\Delta I| \ll (1-{\cal T}) I_0
        \sim S_I/e$.

\bibitem{obviously} Actually, this statement is not so trivial but
        follows from the general quantum mechanical ideas. 

\bibitem{strong} If we formally apply a similar approach to 
	the case  $\Delta I \sim I_0$, we obtain  
	$\langle \mbox{Re}\sigma_{12}(\tau )\rangle /(\sigma_{11}(0)
	\sigma_{22}(0))^{1/2} \leq [2\sqrt{S_{I1} S_{I2}}/
	(S_{I1}+S_{I2})]^{1/2} \exp [-\tau (\Delta I)^2 /
	2(S_{I1}+S_{I2})] $. Then the result of Ref.\ \cite{Gurvitz}
	is smaller than our upper bound for large $\tau $, while
	it is larger for $\tau \alt \Gamma_d^{-1} \sim e/I_0$.
 	The latter unphysical
	situation is because the detector cannot be described classically
	in this case.

\bibitem{alternat} Notice that Eqs.\ (\protect\ref{s1-2}) and 
        (\protect\ref{s_12-g}) can be readily obtained from the
        standard ``reduction'' procedure, $\sigma (\tau ) = 
        {\cal A}/\mbox{Tr} {\cal A}$, ${\cal A} =
        {\cal P} (\langle I\rangle ,\tau ) \, \sigma (0) \, 
        {\cal P} (\langle I\rangle ,\tau )$, if the generalized
        ``projection'' operator ${\cal P} (\langle I\rangle ,\tau )$
        is defined as ${\cal P}_{ii}=[P_i(\langle I\rangle ,\tau )]^{1/2}$,
        ${\cal P}_{12}={\cal P}_{21}=0$.

\bibitem{Misra} B. Misra and E. C. G. Sudarshan, J. Math. Phys.
        {\bf 18}, 756 (1977).

\bibitem{move} To move the electron to the first dot with
	certainty (provided the pure wavefunction) one can,
	for example, 
	create $\epsilon = [(1-4|\sigma_{12}|^2)^{1/2}-1] 
	H \mbox{Re}\sigma_{12}/ |\sigma_{12}|^{2}$
	and wait for a time $\Delta t =[\pi-\arcsin 
	(\mbox{Im} \sigma_{12} \, \hbar\Omega/H)]/\Omega$.

 
\end{references}
\end{document}